\documentclass[11pt]{article}
\usepackage{amsmath,amssymb,color,graphics,epsfig,cite}
\usepackage{graphicx,subfigure}

\usepackage{amsfonts}

\newcommand{\hoch}[1]{$\, ^{#1}$}

\newcommand{\be}{\begin{equation}}
	\newcommand{\ee}{\end{equation}}
\newcommand{\bea}{\setlength\arraycolsep{2pt} \begin{eqnarray}}
	\newcommand{\eea}{\end{eqnarray}}
\newcommand{\nn}{\nonumber}

\def\ft#1#2{{\textstyle{\frac{\scriptstyle #1}{\scriptstyle #2} } }}
\def\fft#1#2{{\frac{#1}{#2}}}

\def\0{{\sst{(0)}}}
\def\1{{\sst{(1)}}}
\def\2{{\sst{(2)}}}
\def\3{{\sst{(3)}}}
\def\4{{\sst{(4)}}}
\def\5{{\sst{(5)}}}
\def\6{{\sst{(6)}}}
\def\7{{\sst{(7)}}}
\def\8{{\sst{(8)}}}
\def\sst#1{{\scriptscriptstyle #1}}

\begin{document}
	
\begin{center}
		{\Large {\bf Black Hole Mass/Charge Relation and \\ Weak No-hair Theorem Conjecture}}
		
		\vspace{20pt}
		
	Guan-Yi Lu\hoch{1}, Meng-Nan Yang\hoch{1} and H. L\"u\hoch{2,1,3}
		
		\vspace{10pt}

		{\it \hoch{1}The International Joint Institute of Tianjin University, Fuzhou,\\ Tianjin University, Tianjin 300350, China}

\medskip

		{\it \hoch{2}Center for Joint Quantum Studies, Department of Physics,\\
			School of Science, Tianjin University, Tianjin 300350, China }

\medskip

{\it \hoch{3}Peng Huanwu Center for Fundamental Theory, Hefei, Anhui 230026, China}
		
		\vspace{40pt}
		
		\underline{ABSTRACT}

	\end{center}

We consider Einstein gravity minimally coupled to two Maxwell fields and one (real) dilaton scalar. We study the electrically-charged spherically-symmetric and static solutions that are asymptotic to Minkowski spacetime. General solutions are described by four independent parameters: mass $M$, two electric charges $(Q_1,Q_2)$ and the scalar charge $\Sigma$. For black holes, the scalar charge is not independent, but a function of $(M,Q_1,Q_2)$. We provide a set of formulae relating the mass to the charges, which allows us to determine $\Sigma(M,Q_1,Q_2)$ without having to solve the black hole equations. Our results confirm the weaker version of the no-hair theorem conjecture involving one real scalar: it can be turned on in a black hole, but it does not have a continuous independent hairy parameter.

\vfill {\footnotesize lgy20020926@tju.edu.cn \ \ \ yangmengnan@tju.edu.cn\ \ \ mrhonglu@gmail.com (corresponding author)}
	
	\thispagestyle{empty}
	\pagebreak
	
	

\section{Introduction}
\label{sec:introduction}

This work involves two related motivations. The first is to construct and study four-dimensional charged black holes in Einstein-Maxwell-Maxwell-Dilaton (EMMD) gravity. The theory is inspired by string theory, whose low-energy effective theory in four dimensions is Einstein gravity minimally coupled to a set of Maxwell fields, together with a scalar coset involving dilaton and axion scalars. One of the simpler and nontrivial supergravity models for studying black hole physics is the ${\cal N}=2$ STU supergravity model involving four Maxwell fields and three $SL(2,\mathbb R)/SO(2)$ scalar cosets \cite{Duff:1995sm}. The three axions can be set to zero in the purely electrically-charged spacetime. The theory can be further truncated to Einstein-Maxwell (EM) or Einstein-Maxwell-Dilaton (EMD) with dilaton coupling constant $a=\sqrt3, 1, 1/\sqrt3,0$ \cite{Duff:1994jr,Gibbons:1994vm,Lu:1995yn}. (The case of $a=\sqrt3$ corresponds to Kaluza-Klein theory and $a=0$ gives the EM theory.) Charged static and rotating black holes in supergravity and gauged supergravity have been constructed \cite{Duff:1999gh,Sabra:1999ux,
Cvetic:1996kv,Chong:2004na}. The static black holes are parameterized by the mass and electric or magnetic charges. Although the dilatons are necessarily turned on, they do not have any continuous independent hairy parameter. In the extremal limit, the mass becomes a function of the Maxwell charges only. There are three classes of mass/charge relations. Considering two-charge examples, they are
\bea
\hbox{subadditivity:}&&\qquad M(Q_1,Q_2)< M(Q_1) + M(Q_2)\,,\label{subadd}\\
\hbox{linear additivity:}&&\qquad M(Q_1,Q_2)= M(Q_1) + M(Q_2)\,,\label{linear}\\
\hbox{super-additivity:}&&\qquad M(Q_1,Q_2)> M(Q_1) + M(Q_2)\,.\label{supadd}
\eea
An example for subadditivity \eqref{subadd} is provided by the extremal dyonic Reissner-Nordstr\"om (RN) black hole, with $M=\sqrt{Q^2 + P^2}$; examples for linear additivity are provided by the supersymmetric black holes in STU supergravity where the black holes are bound states with threshold binding energy \cite{Rahmfeld:1995fm}. A super-additivity example is the extremal dyonic Kaluza-Klein black hole, where $M=(Q^{\fft23} + P^{\fft23})^{\fft32}$ \cite{Gibbons:1987ps,Gibbons:1985ac}. In \cite{Poletti:1995yq, Geng:2018jck, Cribiori:2022cho, Cremonini:2023vwf,Cremonini:2024eog}, the mass/charge relation in extremal dyonic black holes for a general class of EMD theory beyond string theories was studied. In this paper we extend the discussion to consider EMMD theories. Our discussion covers the situations in the EMD theories. The general STU supergravity model can be reduced to special EMMD theories when pairwise Maxwell fields are set equal, or three Maxwell fields are set equal. We will show the mass/charge additivity property depends on the dilaton coupling constants. Furthermore, we shall extend the discussion to non-extremal black holes as well.

The second motivation of this paper is to address the no-hair theorem associated with a scalar field, as such a scalar is necessarily turned on in black holes of the EMMD theory. The no-hair theorem implies that asymptotically-flat, spherically-symmetric and static black holes are specified by the mass and the Maxwell charges only. A stronger version of the no-hair theorem excludes the scalar hair altogether. This can be easily established for free scalars, or a scalar with convex potential. Black holes in Starobinsky model were also shown to carry no hair associated with the scalar mode \cite{Nelson:2010ig,Lu:2015cqa}. However, counterexamples can be easily constructed. Einstein gravity with higher-derivative extensions can admit black holes carrying massive spin-2 hair \cite{Lu:2015cqa,Huang:2022urr,Li:2025dpo}. Scalars with suitable concave potential can support a black hole, e.g.,~\cite{Anabalon:2012ta,Anabalon:2013qua, Gonzalez:2013aca,Feng:2013tza}. Charged black holes in EMD theories can carry scalar hair when the scalar is non-minimally coupled to the Maxwell field \cite{Gibbons:1987ps,Garfinkle:1990qj}.  Black hole scalarization can also be induced by spacetime curvature \cite{Doneva:2017bvd, Antoniou:2017acq,Silva:2017uqg,Cunha:2019dwb}. The mechanism allows to turn on the Starobinsky scalar mode \cite{Liu:2020yqa}.

However, there is one common feature in all these examples of the ``evasion'' of the no-hair theorem, namely, none of the scalar hair parameters in the black holes is a continuous independent parameter. They are all fixed by the mass and/or Maxwell charges. By contrast, in a general solution, a scalar has a nontrivial free parameter, arising as an integration constant from its second-order differential equation. The former is called secondary hair and the latter is called primary hair in literature. (See a review paper on the subject \cite{Herdeiro:2015waa}.) This leads to a weaker version of the black hole no-hair theorem conjecture involving one real scalar: black holes are specified by the mass and (Maxwell) charges only; the scalar can be turned on in a black hole solution, but it has no independent parameters. Note that we do not consider rotations in this paper; however, we would like to give some brief comments on the construction of continuous primary scalar hairy kerr black holes \cite{Herdeiro:2014goa}. The theory involves a complex scalar, in which case, the most general solution should contain one complex or two real scalar hair parameters, in addition to the mass and angular momentum. However, the hairy Kerr black hole carries just one fine-tuned continuous (primary) scalar hair. Another analogous example is provided by the Einstein-Weyl-Maxwell scalar theory, where the most general spherically-symmetric and static solutions contain both the scalar hair and massive spin-2 hair; however, the hairy black hole contains just one additional fine-tuned continuous hair besides the mass and electric charge \cite{Zou:2020rlv}. Thus, for the more general cases, the weak no-hair theorem conjecture should be stated as follows: when non-gauge fields are involved, the most general hairy solution is not a black hole, which requires a further fine-tuning of these continuous hairy parameters. In our case, the theory involves only one real scalar, then the scalar hair must be secondary, fixed by the mass and charges.

It is a formidable task for us to prove such a general conjecture. We shall approach this with a special class of theories. A concrete, explicit example is provided by the EMD theories. Intuitively, if there were a black hole with independent scalar hair parameter, it would continue to be a black hole even after we turn off the Maxwell charge. This would violate the stronger version of the no-hair theory associated with a free massless scalar. However, this argument does not constitute a proof since it assumes that there is a smooth neutral limit in the general solution. It turns out that electrically-charged black hole solutions with two independent parameters, $(M,Q)$, were constructed in \cite{Gibbons:1987ps,Garfinkle:1990qj}. However, these are not the most general asymptotically-flat, spherically-symmetric and static charged solutions. The most general ones involve an additional independent scalar hair parameter \cite{Lu:1996hh}. These three-parameter solutions generally have a naked singularity unless the scalar hair parameter is fine-tuned to be a specific function of $(M,Q)$, thereby confirming the weak no-hair theorem conjecture.

In this paper, we shall study the weak no-hair theorem conjecture in the context of the more complicated EMMD theory. Typically, the scalar hair parameters are not conserved charges. However, in the EMMD theory, it has a global dilaton shifting symmetry, compensated by the appropriate constant scaling of the Maxwell potential fields. This implies that we can describe the scalar hair parameter as a sensible scalar charge. The general solutions contain an independent scalar charge $\Sigma$, in addition to the mass $M$, electric charges $(Q_1,Q_2)$ of the two Maxwell fields. However, for the solution to describe a black hole, the scalar charge $\Sigma$ has to be fine-tuned to be a specific function of mass and charges, i.e., $\Sigma=\Sigma(M,Q_1,Q_2)$. Since there are no general exact solutions in a generic EMMD theory, we would have to establish this equation by numerical methods, which is generally inefficient since numerical methods can only be performed on a case-by-case basis. In this paper, based on special examples and arguments, we propose a set of algebraic and partial differential equations for the mass $M$, electric charges $(Q_1,Q_2)$ and the scalar charge $\Sigma$ that black holes should satisfy. These equations enable us to determine $\Sigma(M,Q_1,Q_2)$ for a black hole without having to solve the black hole equations of motion. This provides a quicker way to confirm the weak no-hair theorem conjecture.

The paper is organized as follows. In section 2, we introduce the general EMMD theory. We present the spherically-symmetric and static ansatz for black holes charged under two Maxwell fields. We obtain the equations of motion and analyse both the asymptotic and horizon geometries. General solutions contain a free scalar charge parameter $\Sigma$, in addition to $(M,Q_1,Q_2)$. However, special solutions and sample numerical analysis for the general case all indicate that $\Sigma$ is not independent, but it must be a fine-tuned function of $(M,Q_1,Q_2)$. In section 3, we review the known examples of exact black hole solutions, which enable us to propose in section 4 some algebraic and differential relations between $(M,Q_1,Q_2)$ and $\Sigma$. This allows us to determine the black hole scalar hair charge $\Sigma$ as a function of $(M,Q_1,Q_2)$. Some sample numerical tests show these parameters indeed specify a black hole, confirming the weak no-hair theorem conjecture. We conclude the paper in section 5.

\section{Charged black holes in EMMD theories}
\label{sec:emmdbh}

\subsection{The theory}

We begin with a brief introduction of Einstein-Maxwell-Maxwell-Dilaton (EMMD) theories in four dimensions. It is Einstein gravity minimally coupled to two Maxwell fields $(A_1,A_2)$ and one (real) dilaton scalar $\phi$. The dilaton is non-minimally coupled to the Maxwell fields with exponential couplings. The Lagrangian takes the form
\be
{\cal L}=\sqrt{-g} \Big(R - \ft12 (\partial \phi)^2 - \ft14 e^{a_1\phi} (F_1)^2 - \ft14 e^{a_2 \phi} (F_2)^2\Big)\,,\qquad F_1=dA_1\,,\qquad F_2=dA_2\,.\label{genlag}
\ee
The dilaton is massless, and the Maxwell-dilaton couplings are specified by the two dilaton coupling constants $(a_1,a_2)$. The Lagrangian is invariant under the dilatations generated by a constant shift of the dilaton
\be
\phi\rightarrow \phi + c\,,\qquad A_i \rightarrow A_i\, e^{-\fft12a_i\,c}\,,\qquad i=1,2.
\label{dilatonscale}
\ee
Furthermore, the equations of motion are invariant under the constant trombone scaling \cite{Cremmer:1997xj}
\be
g_{\mu\nu} \rightarrow \Omega^2 g_{\mu\nu}\,,\qquad
A_i \rightarrow \Omega\, A_i\,.\label{trombonescale}
\ee
These global symmetries, together with the local gauge symmetry of the Maxwell fields, ensure that we can always write the Minkowski vacuum in spherical polar coordinates in the standard form, namely
\be
ds^2 = -dt^2 + dr^2 + r^2 d\Omega_2^2\,,\qquad \phi=0\,,\qquad A_i=0\,.\label{vacuum}
\ee
Note that we can always perform a Hodge dual on $A_1$ and/or $A_2$ fields, such that the respective dilaton coupling constants $a_i$ change their signs. Thus, we can set without loss of generality that
\be
a_1 >0\,,\qquad a_2<0\,.
\ee
In this case, the dilaton can be decoupled, i.e., $\phi=0$, provided that $a_1 F_1^2=- a_2 F_2^2$. Redefining the gauge fields as
\be
A_1=\sqrt{\fft{-a_2}{a_1-a_2}}\, A\,,\qquad A_2 = \sqrt{\fft{a_1}{a_1-a_2}}\, A\,,\label{truncation}
\ee
the Lagrangian \eqref{genlag} then consistently reduces to Einstein-Maxwell gravity
\be
{\cal L}=\sqrt{-g} (R - \ft14 F^2)\,,\qquad F=dA\,.
\ee
Note that the dilaton symmetry \eqref{dilatonscale} also implies that if we redefine the gauge fields as
\be
A_1\rightarrow e^{-\fft12a_1 \phi} A_1\,,\qquad A_2\rightarrow e^{-\fft12a_2 \phi} A_2\,,\label{newA}
\ee
the dilaton $\phi$ enters the Lagrangian only through a derivative $\partial_\mu \phi$. In this choice of fields, the scalar equation of motion can be expressed as
\be
\nabla_\mu J^\mu (\partial \phi,A_i)=0\,.\label{dilatoncons}
\ee
This implies that we can define a scalar charge, as in the case of electric charges of Maxwell theory, to describe the scalar hairy parameter. Of course, in the new $A_i$ definitions, the bare gauge potential will appear in the Lagrangian, and hence the gauge symmetry becomes non-manifest.
The dilaton conservation law \eqref{dilatoncons} also implies the existence of a radially-independent quantity for stationary solutions. Thus the scalar charge can be related to the horizon data, as was observed in \cite{Pacilio:2018gom,Herdeiro:2025blx}.

However, it should be emphasized that for the purpose of this paper, it is inessential whether the scalar hair parameter is a charge or just a parameter of integration constant. It is only simpler lexically to describe the parameter as the scalar charge.

\subsection{Black hole ansatz and equations of motion}

In this paper, we consider spherically-symmetric and static black holes, electrically charged by both the Maxwell fields. The general ansatz using the Schwarzschild-like radial coordinate takes the form
\be
ds^2 = - h dt^2 + \fft{dr^2}{f} + r^2 d\Omega_2^2\,,\qquad A_1 = \phi_1(r) dt\,,\qquad
A_2=\phi_2(r) dt\,,\qquad \phi=\phi(r)\,.
\ee
When a scalar is involved, it is not helpful to use the Schwarzschild-like radial coordinate. Following the strategy of \cite{Geng:2018jck, Cribiori:2022cho,Cremonini:2023vwf, Cremonini:2024eog}, we find that a more convenient ansatz is
\bea
ds^2 &=& - H_1^{-\fft2{1+a_1^2}} H_2^{-\fft2{1+a_2^2}} f dt^2 +
H_1^{\fft2{1+a_1^2}} H_2^{\fft2{1+a_2^2}}\Big(\fft{dr^2}{f} + r^2 d\Omega_2^2\Big)\,,\nn\\
F_1 &=&  \fft{4Q_1}{r^2 H_1^2} H_2^{-\fft{2(1+a_1 a_2)}{1+a_2^2}} dt\wedge dr\,,\qquad
F_2 = \fft{4Q_2}{r^2 H_2^2} H_1^{-\fft{2(1+a_1 a_2)}{1+a_1^2}} dt\wedge dr\,,\nn\\
\phi &=& \fft{2a_1}{1 + a_1^2} \log H_1 + \fft{2 a_2}{1 + a_2^2} \log H_2\,,\qquad
f=1 - \fft{2\mu}{r}\,.\label{ansatz}
\eea
Here $H_1$ and $H_2$ are functions of the radial $r$. Note that the radius of the $S^2$ in this radial coordinate gauge is
\be
\rho=r H_1^{\fft1{1+a_1^2}} H_2^{\fft1{1+a_2^2}}\,.
\ee
This general ansatz is motivated by the known singly-charged black holes \cite{Duff:1996hp, Cvetic:1996gq}, corresponding to setting $Q_1=0$, $H_1=1$ or $Q_2=0$, $H_2=1$. Thus, the ansatz is the composition of the two singly-charged ones. The Maxwell equations of the two gauge fields are automatically satisfied. The remaining scalar equation and Einstein field equations reduce to two second-order differential equations of $(H_1,H_2)$:
\bea
0&=&(a_1-a_2)\fft{H_1''}{H_1} + (\ft12 a_2 -a_1) \fft{H_1'^2}{H_1^2}
-\frac{a_2\left(a_1^2+1\right)}{2 \left(a_2^2+1\right)} \fft{H_2'^2}{H_2^2}
-\frac{a_2 \left(a_1 a_2+1\right)}{a_2^2+1} \fft{H_1' H_2'}{H_1 H_2}\nn\\
&& + \frac{2 a_1 (r-\mu )-a_2 (2 r-3 \mu )}{r (r-2\mu )} \fft{H_1'}{H_1} +
\frac{\left(a_1^2+1\right) a_2 \mu }{ \left(a_2^2+1\right) r (r-2\mu )} \fft{H_2'}{H_2}\nn\\
&&+ \frac{2 \left(a_1^2+1\right)}{r^3 (r-2\mu )}\Big(\frac{a_2 Q_2^2 H_1^{-\frac{2 \left(a_1 a_2+1\right)}{a_1^2+1}}}{H_2^2}+\frac{\left(2 a_1-a_2\right) Q_1^2 H_2^{-\frac{2 \left(a_1 a_2+1\right)}{a_2^2+1}}}{H_1^2}\Big)\,,\nn\\
0&=& (a_2 -a_1) \fft{H_2''}{H_2} + (\ft12 a_1-a_2) \fft{H_2'^2}{H_2} - \fft{a_1 (1+a_2^2)}{2(1+a_1^2)}
\fft{H_1'^2}{H_1^2} - \frac{a_1 \left(a_1 a_2+1\right)}{a_1^2+1} \fft{H_1' H_2'}{H_1 H_2}\nn\\
&&+\frac{2 a_2 (r-\mu )-a_1 (2 r-3 \mu )}{r (r-2\mu )} \fft{H_2'}{H_2} +
\frac{a_1 \left(a_2^2+1\right) \mu }{\left(a_1^2+1\right) r (r-2\mu )}\fft{H_1'}{H_1}\nn\\
&&+ \frac{2 \left(a_2^2+1\right)}{r^3 (r-2\mu )}\Big(
\frac{a_1 Q_1^2 H_2^{-\frac{2 \left(a_1 a_2+1\right)}{a_2^2+1}}}{H_1^2}+\frac{\left(2 a_2-a_1\right) Q_2^2 H_1^{-\frac{2 \left(a_1 a_2+1\right)}{a_1^2+1}}}{H_2^2}\Big)\,,\label{eom2}
\eea
together with their first-order Hamiltonian constraint
\bea
0&=& \fft{1}{1+a_1^2} \fft{H_1'^2}{H_1^2} + \fft{1}{1+a_2^2} \fft{H_2'^2}{H_2^2} + \frac{2 \left(a_1 a_2+1\right)}{\left(a_1^2+1\right) \left(a_2^2+1\right)} \fft{H_1'H_2'}{H_1 H_2}\nn\\
&&
-\fft{2\mu}{r(r-2\mu)} \Big( \fft{1}{1+a_1^2} \fft{H_1'}{H_1} + \fft{1}{1+a_2^2} \fft{H_2'}{H_2}\Big)\nn\\
&&-\fft{4}{r^3 (r-2\mu)}\Big(\frac{Q_2^2 H_1^{-\frac{2 \left(a_1 a_2+1\right)}{a_1^2+1}}}{H_2^2}+\frac{Q_1^2 H_2^{-\frac{2 \left(a_1 a_2+1\right)}{a_2^2+1}}}{H_1^2}\Big)\,.
\eea
Note that we can use the global symmetries \eqref{dilatonscale} and \eqref{trombonescale} to choose the integration constants at asymptotic infinity such that
\be
H_1(\infty)=1=H_2(\infty)\,,\qquad \phi(\infty)=0\,.\label{infinityHphi}
\ee
The consequent metric at the asymptotic infinity is the Minkowski vacuum \eqref{vacuum}. The two parameters $(Q_1,Q_2)$ are simply the electric charges of the Maxwell fields $(A_1,A_2)$ respectively. The definition of the charges is
\be
Q_i = \fft{1}{16\pi} \int e^{a_i \phi} {*F_i}\,.\label{Qcharges}
\ee
Note that when $a_2=-a_1$, the black hole with two electric charges is equivalent to the dyonic black hole of the EMD theory involving only the $A_1$ field, studied in \cite{Poletti:1995yq, Geng:2018jck, Cribiori:2022cho, Cremonini:2023vwf,Cremonini:2024eog}.

\subsection{Asymptotic infinity}

For general dilaton coupling constants $(a_1,a_2)$, there are no analytic charged black hole solutions with two independent electric charges $(Q_1,Q_2)$. Charged black hole solutions exist for some special values of $(a_1,a_2)$ and we shall discuss them in the next section. In this paper, our focus is not to construct new exact charged black hole solutions in the EMMD theory; rather, we shall try to establish the weak no-hair theorem within the EMMD theory that spherically-symmetric and static black holes are parameterized by the mass and electric charges $(M,Q_1,Q_2)$ only. Although the dilaton scalar is necessarily turned on in these black holes, it does not have its own independent scalar charge. This statement is not at all apparent, even for such a simple theory. The equations of motion \eqref{eom2} and their Hamiltonian constraint imply that there are a total of three additional integration constants, beyond the already existing mass/charge parameters $(\mu, Q_1,Q_2)$.

It is clear that not all the parameters will lead to the asymptotic Minkowski region or have the horizon structure. We first examine the asymptotic large-$r$ expansion of $H_i$, with boundary condition \eqref{infinityHphi}. For low-lying falloffs, we find
\bea
H_1 &=&  1 + (1 + a_1^2) \Big(\fft{m_1}{r} + \fft{n_2+m_2}{r^2} + \fft{m_3}{r^3} + \cdots\Big),\nn\\
H_2 &=&  1 + (1 + a_2^2) \Big(\fft{n_1}{r} + \fft{n_2-m_2}{r^2} + \fft{n_3}{r^3} + \cdots\Big),
\label{larger}
\eea
where
\bea
Q_1 &=& \frac{1}{2} \sqrt{m_1 \left(\left(a_1^2+1\right) m_1+\left(a_1 a_2+1\right) n_1+2 \mu \right)-2 m_2}\,,\nn\\
Q_2 &=& \frac{1}{2} \sqrt{n_1 \left(\left(a_1 a_2+1\right) m_1+\left(a_2^2+1\right) n_1+2 \mu \right)+2 m_2}\,,\nn\\
n_2 &=& -\frac{1}{2} \left(a_1 a_2+1\right) m_1 n_1\,,\nn\\
m_3 &=& \frac{1}{6} \left(2 \left(a_1 a_2+1\right){}^2 m_1 n_1^2+\left(a_1 a_2+1\right) n_1 \left(\left(a_1^2+1\right) m_1^2-4 m_2\right)+2 m_2 \left(\left(a_1^2+1\right) m_1+4 \mu \right)\right),\nn\\
n_3 &=& \frac{1}{6} \left(2 \left(a_1 a_2+1\right)^2 m_1^2 n_1+\left(a_1 a_2+1\right) m_1 \left(\left(a_2^2+1\right) n_1^2+4 m_2\right) -2 m_2 \left(\left(a_2^2+1\right) n_1+4 \mu \right)\right).\nn
\eea
Thus, we see that the asymptotic expansion is parameterized by four parameters, namely
$(m_1,n_1,m_2,\mu)$, with the electric charges $(Q_1,Q_2)$ given above. All the higher mode coefficients can be expressed in terms of these four parameters. The mass of the spacetime configuration can be read off from $g_{tt}=-1 + 2M/r + \cdots$, given by
\be
M=m_1 + n_1 + \mu\,.
\ee
The (massless) scalar charge $\Sigma$ is defined by
\be
\phi = \fft{4\Sigma}{r} + \cdots\,,\qquad \Sigma=\ft12 (a_1 m_1 + a_2 n_1)\,.
\ee
We see that the asymptotic Minkowski geometry is specified by four {\it independent} parameters, related to the four hairs, the mass $M$, two electric charges $(Q_1,Q_2)$ and the scalar charge $\Sigma$, all of which can be expressed as functions of $(m_1,n_1,m_2,\mu)$. However, for a generic choice of the four parameters, the solution does not describe a black hole. Instead it has generally naked singularities. This conclusion can be straightforwardly confirmed by a numerical approach of some sample examples. We can select a generic value of $(m_1,n_1,m_2,\mu)$ such that $(Q_1,Q_2)$ are real and $M>0$.  We then use the above asymptotic expansion for sufficiently large $r$ as the boundary and integrate the equations inward to the middle of the spacetime. We find that the functions $(H_1,H_2)$ will inevitably diverge or vanish on the horizon $r_+=2\mu$. However, charged black holes do exist. For a given choice $(Q_1,Q_2,M)$, or more conveniently $(Q_1,Q_2,\mu)$, we can fine-tune the parameters $(m_1,n_1,m_2)$ so that a numerical solution of a charged black hole of given mass and charges is constructed. This evidence indicates that black hole solutions are specified by the mass and Maxwell charges, while the scalar charge $\Sigma$ is not independent, but a function of $(M,Q_1,Q_2)$. This type of numerical approach, however, is cumbersome and works only on a case-by-case basis, and therefore it is not worth the effort to map out the function $\Sigma(M,Q_1,Q_2)$ for general $(a_1,a_2)$ in this approach.

We now study the long-range force between two identical spacetime configurations. For spherically-symmetric and static configurations, separated at large distance, the long-range force is given by \cite{Cremonini:2022sxf}
\be
\lim_{r\rightarrow \infty} r^2\, \hbox{Force} = 4 Q_1^2 + 4 Q_2^2 - M^2 - 4\Sigma^2\,.
\ee
Note that the signs above are consistent with the standard quantum field theory where the mutual force between two identical particles mediated by a spin-1 field is universally repulsive and the forces mediated by spin-2 or spin-0 fields are universally attractive. The factor 4 arises from our convention of the kinetic term coefficients of the Maxwell fields. In \cite{Cremonini:2022sxf} and subsequent later works \cite{Cremonini:2023vwf,Cremonini:2024eog}, extremal black holes were considered, in which case, the long-range force vanishes. We would like consider to more general case. It follows from our ansatz and its large-$r$ expansion \eqref{larger} that the long-range force is universally given by
\be
4 Q_1^2 + 4 Q_2^2 - M^2 - 4\Sigma^2 = -\mu^2\,.\label{longrange}
\ee
It is important to note that the long-range force has a universal relation for general parameters $(m_1,n_1,m_2)$. This is understandable since the long-range force is a local concept and it should not depend on whether the spacetime configuration is a black hole or not. The universal nature of this long-range force formula \eqref{longrange} indicates that the parameter $\mu$ specifies the overall attractive force, which can be a priori introduced. It is somewhat intriguing that it arises naturally from our ansatz, as the parameter specifying the horizon radius. In the extremal limit $\mu=0$, two identical black holes experience no force at long range. In supergravities, these types of extremal black holes are also supersymmetric, in which case, the mass/charge relation is linear and the no-force condition can apply in all ranges of separations.

\subsection{Near-horizon geometry}

Having established that the asymptotic structure has an extra parameter beyond the black hole configuration, it is natural to ask what happens on the horizon at $r_+=2\mu$. If the horizon configuration has only three parameters total, then we can establish that the scalar charge is not an independent parameter. We assume that functions $(H_1,H_2)$ are analytic near the horizon $r_+$. We can perform Taylor expansions and we find
\bea
H_1 &=& h_{10} + h_{11} (r-2\mu) + h_{12} (r-2\mu)^2 + \cdots\,,\nn\\
H_2 &=& h_{20} + h_{21} (r-2\mu) + h_{22} (r-2\mu)^2 + \cdots\,,
\eea
where
\bea
h_{11}&=&-\frac{\left(a_1^2+1\right) Q_1^2 h_{20}^{-\frac{2 \left(a_1 a_2+1\right)}{a_2^2+1}}}{2 \mu ^3 h_{10}}\,,\qquad h_{21}=-\frac{\left(a_2^2+1\right) Q_2^2 h_{10}^{-\frac{2 \left(a_1 a_2+1\right)}{a_1^2+1}}}{2 \mu ^3 h_{20}}\,,\nn\\
h_{12} &=& \frac{a_1^2+1}{8 \mu ^6} Q_1^2 h_{10}^{-\frac{a_1^2+2 a_2 a_1+3}{a_1^2+1}} h_{20}^{-\frac{2 \left(a_2^2+a_1 a_2+2\right)}{a_2^2+1}} \left(2\mu ^2 h_{20}^2 h_{10}^{\frac{2( a_1 a_2+1)}{a_1^2+1}}-\left(a_1 a_2+1\right) Q_2^2\right),\nn\\
h_{22} &=& \frac{a_2^2+1}{8 \mu ^6} Q_2^2 h_{10}^{-\frac{2 \left(a_1^2+a_2 a_1+2\right)}{a_1^2+1}} h_{20}^{-\frac{a_2^2+2 a_1 a_2+3}{a_2^2+1}} \left(2\mu ^2 h_{10}^2 h_{20}^{\frac{2( a_1 a_2+1)}{a_2^2+1}}-\left(a_1 a_2+1\right) Q_1^2\right),
\eea
{\it etc.} The near-horizon geometry is specified by not only the parameters $(\mu,Q_1,Q_2)$, but also two additional parameters $(h_{10}, h_{20})$. The remaining coefficients in the Taylor expansion are functions of all these five parameters. Thus, the near-horizon geometry is not restrictive enough to fix the total number of independent parameters of the charged black hole.

We can use a numerical approach to see that not all the parameters can integrate to large $r$ to give asymptotic Minkowski spacetime. In fact, for a given choice $(Q_1,Q_2,\mu)$, our numerical analysis indicates that a generic value of $(h_{10},h_{20})$ will lead to vanishing $(H_1,H_2)$, which corresponds to a naked curvature singularity. We have to fine-tune two parameters $(h_{10},h_{20})$ to construct a numerical black hole solution, integrating from horizon to asymptotic infinity. This is a much more difficult task compared to the integration from large $r$ to horizon, since we only need to fine-tune one parameter in this latter case. Note that the apparent extra parameter on the horizon is that, after assuming $\phi(\infty)=0$, the scalar on the horizon no longer vanishes.

We are thus in an awkward situation. A careful and tedious numerical calculation of any sample examples can confirm the weak no-hair theorem conjecture, but a proof is absent. In the next section, we shall improve our understanding by examining the known special examples of exact black hole solutions.

Before continuing, we note that although the temperature and entropy requires specific knowledge of the black hole solution, given by
\be
T=\fft{1}{8\pi \mu} H_1^{-\fft{2}{1+a_1^2}} H_2^{-\fft{2}{1+a_2^2}}\Big|_{r=2\mu}\,,\qquad
S=4\pi\mu^2 H_1^{\fft{2}{1+a_1^2}} H_2^{\fft{2}{1+a_2^2}}\Big|_{r=2\mu}\,,
\ee
their product is universally determined by the net long-range force parameter $\mu$, i.e.
\be
2 T S = \mu\,.\label{tsmu}
\ee
This relates the property of asymptotic region to that of the horizon region. It shows explicitly that the long-range force vanishes at the extremal $(T=0)$ limit. The relation \eqref{tsmu} and the universality of \eqref{longrange} suggest that the parameter $\mu$ is a sensible choice to parameterize a black hole.

\section{Exact special solutions of charged black holes}
\label{sec:specialbh}

In the previous section, we considered charged black holes in EMMD theories. For general dilatonic coupling constants $(a_1,a_2)$, we know no exact black hole solutions with independent electric charges. The analysis on the asymptotic infinity and the near-horizon regions indicates that there are additional parameters beyond the mass and Maxwell charges. We have to consider numerical analysis to see that black holes carry only three hairs $(M,Q_1,Q_2)$. The scalar, while it is necessarily turned on, does not have its own independent charge. The shortcoming of the numerical analysis is that it involves many parameters and we cannot obtain a clear overall picture without conducting a massive survey or scan of the parameter space.  In this section, we analyse the known special solutions of black holes, and try to figure out a general pattern.

\subsection{Case 1: $a_1a_2=-1$}

Exact black hole solutions exist when $a_1 a_2=-1$ \cite{Lu:2013eoa}. We have
\be
H_1=1 + \fft{\sqrt{4(1+a_1^2) Q_1^2+\mu^2} -\mu}{r}\,,\qquad H_2=1 + \fft{\sqrt{4(1+a_1^{-2}) Q_2^2+\mu^2} -\mu}{r}\,.\label{exact1}
\ee
The mass and the scalar charge are
\bea
M &=& \frac{\sqrt{4 \left(a_1^2+1\right) Q_1^2+\mu ^2}}{a_1^2+1} + \frac{\sqrt{4 \left(a_1^{-2}+1\right) Q_2^2+\mu ^2}}{a_1^{-2}+1}\,,\nn\\
\Sigma &=& \frac{a_1 \sqrt{4 \left(a_1^2+1\right) Q_1^2+\mu ^2}}{2 \left(a_1^2+1\right)}
-\frac{a_1^{-1}\sqrt{4 \left(a_1^{-2}+1\right) Q_2^2+\mu ^2}}{2\left(a_1^{-2}+1\right)}\,.\label{masssiga1a2m1}
\eea
It is easy to verify that the hairy parameters $(M,\Sigma, Q_1,Q_2)$ satisfy the long-range force constraint \eqref{longrange}.

By comparing this exact solution to the general large-$r$ expansion, we immediately find that the coefficients $(m_1,n_1,m_2)$ are fully specified by $(Q_1,Q_2,\mu)$ as follows
\be
m_1=\sqrt{4(1+a_1^2) Q_1^2+\mu^2} -\mu\,,\qquad
n_1=\sqrt{4(1+a_1^{-2}) Q_2^2+\mu^2} -\mu\,,\qquad m_2=0\,.
\ee
For $a_1a_2=-1$ case, the most general solution with continuous scalar hair can be constructed \cite{Lu:1996hh}, and it is also easy to demonstrate numerically that for fixed $(Q_1,Q_2,\mu)$, any deviation from the above, say $m_2$ is non-vanishing, the solution ceases to be a black hole. When $m_2$ is non-vanishing, no matter how small it is, either the functions $(H_1,H_2)$ or their derivatives $(H_1',H_2')$ will diverge on the would-be horizon $r=2\mu$.

Alternatively, compared to the horizon expansion, we find that the parameters $(h_{10}, h_{20})$ are also completely fixed, given by
\be
h_{10}=\frac{\sqrt{4 \left(a_1^2+1\right) Q_1^2+\mu ^2}+\mu }{2 \mu }\,,\qquad
h_{20}=\frac{\sqrt{4 \left(a_1^{-2}+1\right) Q_1^2+\mu ^2}+\mu }{2 \mu }\,.
\ee
Again, for fixed mass and charges, any deviation from the above horizon data will cause the functions $(H_1,H_2)$ to be unable to integrate out from $r=2\mu$ to asymptotic infinity to become some finite values.

In fact, the numerical conclusion can also be drawn analytically since for $a_1a_2=-1$, exact solutions with four independent parameters $(M,Q_1,Q_2,\Sigma)$ exist \cite{Lu:1996hh}, and only the restricted mass/charge relation \eqref{masssiga1a2m1} gives a black hole.

\subsection{Case 2: $a_1=-a_2=\sqrt3$}

Exact solution also exists when $a_1=-a_2 = \sqrt{3}$; it is given by
\be
H_1 =1 +\frac{q_1}{r} + \frac{q_1 q_2 \left(2 \mu +q_1\right)}{2\left(4 \mu +q_1+q_2\right) r^2}\,,\qquad
H_2=1+\frac{q_2}{r}+\frac{q_1 q_2 \left(2 \mu +q_2\right)}{2 r^2 \left(4 \mu +q_1+q_2\right)}\,.
\ee
The mass $M$, charges $(Q_1,Q_2)$ and the scalar hair $\Sigma$ are related to the parameters $(\mu,q_1,q_2)$ as follows
\bea
M&=&\frac{1}{4} \left(4 \mu +q_1+q_2\right)\,,\qquad \Sigma=\frac{1}{8} \sqrt{3} \left(q_1-q_2\right)\,,\nn\\
Q_1 &=& \frac{1}{4} \sqrt{\frac{q_1 \left(8 \mu ^2+6 \mu  q_1+q_1^2\right)}{4 \mu +q_1+q_2}}\,,\qquad
Q_2 = \frac{1}{4} \sqrt{\frac{q_2 \left(8 \mu ^2+6 \mu  q_2+q_2^2\right)}{4 \mu +q_1+q_2}}\,.
\eea
It is straightforward to verify that the universal long-range force condition \eqref{longrange} is satisfied. Again, for fixed $(Q_1,Q_2,\mu)$, the parameters $(m_1,n_1,m_2)$ in the asymptotic expansion or the parameters $(h_{10}, h_{20})$ are completely fixed. Our numerical analysis indicates that any deviation from these fixed parameters will cause the solution not to describe a black hole.

In the $\mu=0$ extremal limit, we have
\bea
H_1 &=& 1 +\frac{4 Q_1^{2/3} \sqrt{Q_1^{2/3}+Q_2^{2/3}}}{r}+ \frac{8 Q_1^{4/3} Q_2^{2/3}}{r^2}\,,\nn\\
H_2 &=& 1+\frac{4 Q_2^{2/3} \sqrt{Q_1^{2/3}+Q_2^{2/3}}}{r}+\frac{8 Q_2^{4/3} Q_1^{2/3}}{r^2}\,.
\eea
The mass and scalar hair are now given by
\be
M=\left(Q_1^{2/3}+Q_2^{2/3}\right)^{3/2}\,,\qquad
\Sigma=\frac{1}{2} \sqrt{3} \left(Q_1^{2/3}-Q_2^{2/3}\right) \sqrt{Q_1^{2/3}+Q_2^{2/3}}\,.
\ee
This solution is equivalent to the extremal dyonic Kaluza-Klein black hole \cite{Gibbons:1987ps,Gibbons:1985ac}. Note that exact solutions with four independent parameters $(M,Q_1,Q_2,\Sigma)$ exist \cite{Lu:1996hh,Lu:2013uia} and only the the subset above gives rise to a black hole.

\subsection{Case 3: $Q_2/Q_1=\sqrt{a_1/(-a_2)}$}

Finally, for general $(a_1,a_2)$, it follows from \eqref{truncation} that when the charges are
\be
Q_1 = \sqrt{\fft{-a_2}{a_1-a_2}}\, Q\,,\qquad Q_2 = \sqrt{\fft{a_1}{a_1-a_2}}\, Q\,,
\ee
the solution becomes the RN black hole, with
\be
H_1 = \left(1+\frac{\sqrt{\mu ^2+4 Q^2}-\mu }{r}\right)^{-\frac{a_2\left(a_1^2+1\right) }{a_1-a_2}}\,,\qquad
H_2=\left(1+\frac{\sqrt{\mu ^2+4 Q^2}-\mu }{r}\right)^{\frac{a_1 \left(a_2^2+1\right)}{a_1-a_2}}\,.
\ee
The mass and scalar charge are
\be
M=\sqrt{4Q^2 + \mu}\,,\qquad \Sigma=0\,.
\ee
Again the relation \eqref{longrange} is satisfied.

All these special examples suggest that a black hole solution satisfies the weak no-hair theorem in EMMD theories such that the electrically-charged black holes are specified only by the mass and the conserved Maxwell charges $Q_i$. The scalar charge $\Sigma$, although being nonzero, is not independent, but a function of mass and $Q_i$. We shall formulate the weak no-hair theorem of EMMD theories in the next section.

\section{A weak no-hair theorem conjecture}

In section \ref{sec:emmdbh}, we considered a general class of spherically-symmetric and static  black holes carrying electric charges under two Maxwell fields in general EMMD theories. The analysis of the near-horizon geometry and asymptotic infinity falloffs indicated that there might exist additional independent scalar charges beyond the mass and electric charges of the Maxwell fields. However, both special solutions and numerical analysis showed that the scalar charge is not an independent parameter. In section \ref{sec:specialbh}, we reviewed some known black holes for special values of the dilaton coupling constants. All these suggest that the black holes are specified in terms of their mass and charges $Q_i$ only. In terms of parameters $(Q_1,Q_2,\mu)$, we must have
\be
M=M(Q_1,Q_2,\mu)\,,\qquad \Sigma=\Sigma(Q_1,Q_2,\mu)\,.
\ee
We now propose a set of conditions that these parameters of a black hole must satisfy. The first condition is that these parameters satisfy the universal long-range force formula \eqref{longrange}. Secondly, $M$ and $\Sigma$ must both be homogeneous functions of these variables. This conclusion is based on the dimensional analysis and also the lack of a parameter in the theory to break this homogeneity. Thus, we have
\be
M=Q_1 \fft{\partial M}{\partial Q_1} + Q_2 \fft{\partial M}{\partial Q_2} + \mu \fft{\partial M}{\partial \mu}\,,\qquad
\Sigma=Q_1 \fft{\partial \Sigma}{\partial Q_1} + Q_2 \fft{\partial \Sigma}{\partial Q_2} + \mu \fft{\partial \Sigma}{\partial \mu}\,.\label{homo}
\ee
Finally the dilaton scaling symmetry indicates the following scalar hair equation
\be
\Sigma=\frac12 a_1 Q_1 \fft{\partial M}{\partial Q_1} +
\frac12 a_2 Q_2 \fft{\partial M}{\partial Q_2}+\mu \fft{\partial \Sigma}{\partial\mu}\,.\label{scalarhaireq}
\ee
In the extremal limit $\mu=0$ and $a_1=-a_2$, these mass/charge relations were proposed in \cite{Cremonini:2023vwf,Cremonini:2024eog} in the context of dyonic black holes in EMD theories. Our extension to general $(a_1,a_2)$ and non-extremality, e.g., the last term in \eqref{scalarhaireq}, is based on the various special examples reviewed in the previous section. In the next, we shall illustrate that we can use the above four rules to derive the mass/charge relations without solving the black hole equations of motion.

\subsection{Case 1: $Q_2=0$}

When $Q_2=0$, the Maxwell field strength $F_2$ is turned off, and the theory reduces to EMD gravity. An exact solution of the black hole exists, as given in \eqref{exact1} with $Q_2=0$. In this case, $H_2=1$. It should be pointed out that the function $H_2$ in the general solution is not a constant, even after we turn off $F_2$, but it would cease to describe a black hole. We now demonstrate that the black hole mass/charge relation can be obtained from the above four rules without having to solve the black hole equations of motion or knowing the solution of $H_1$.

The homogeneity relations of \eqref{homo}, after setting $Q_2=0$, imply that
\be
M=\mu f(x)\,,\qquad \Sigma = \mu g(x)\,,\qquad x=\fft{Q_1}{\mu}\,.
\ee
The scalar-hair equation \eqref{scalarhaireq} implies $2g'(x)=a_1 f'(x)$. Thus, we have
\be
g(x) =c+ \ft12 a_1 f(x)\,.
\ee
When $f(x)=1$, we have $M=\mu$, which is the mass of the Schwarzschild black hole. Thus we determine the integration constant $c=-\fft12 a_1$. Substituting all the results into the long-range force formula \eqref{longrange}, we have
\be
f(x) = \frac{\sqrt{4 \left(a_1^2+1\right) x^2+1}+a_1^2}{a_1^2+1}\,.
\ee
We thus have
\be
M=\frac{a_1^2 \mu +\sqrt{4 \left(a_1^2+1\right) Q_1^2+\mu ^2}}{a_1^2+1}\,,
\qquad
\Sigma=\frac{a_1 \left(\sqrt{4 \left(a_1^2+1\right) Q_1^2+\mu ^2}-\mu \right)}{2 \left(a_1^2+1\right)}\,.
\ee
This is precisely the result of \eqref{masssiga1a2m1} after setting $Q_2=0$. Note that it is necessary to impose the boundary condition on \eqref{scalarhaireq}, namely $\Sigma=0$ when $Q_1=Q_2=0$.

\subsection{Case 2: $\mu=0$}

\begin{figure}
	\centering
	\includegraphics[width=0.4\linewidth]{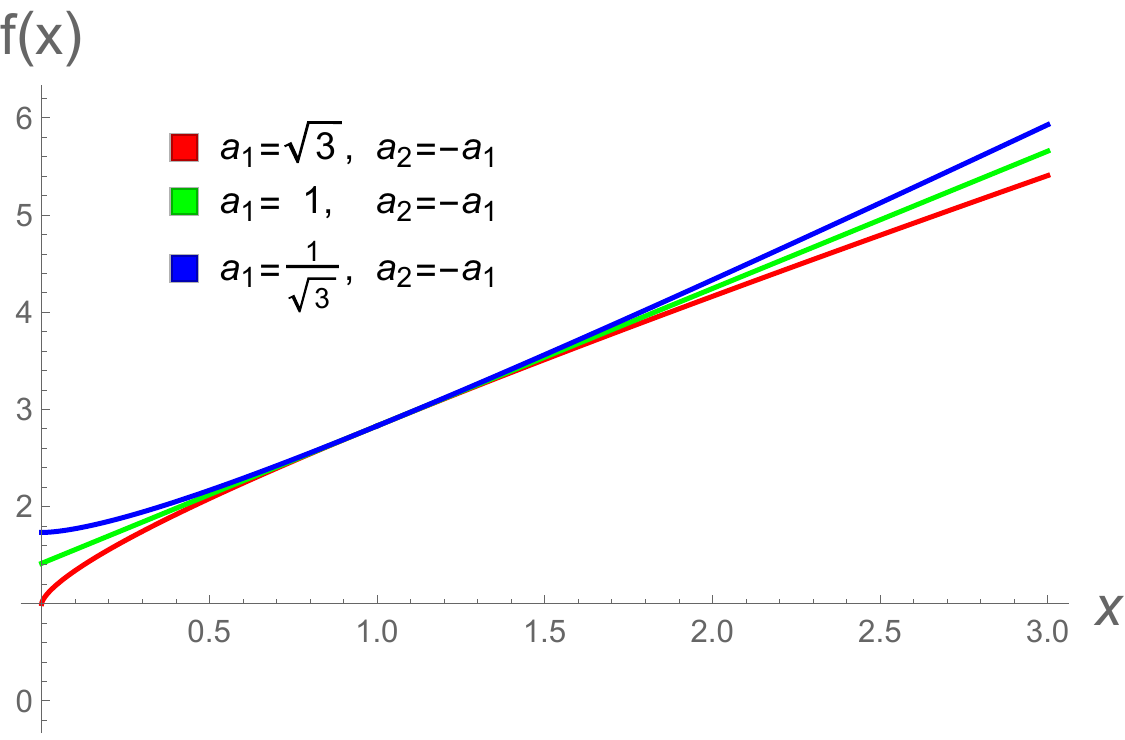} \includegraphics[width=0.4\linewidth]{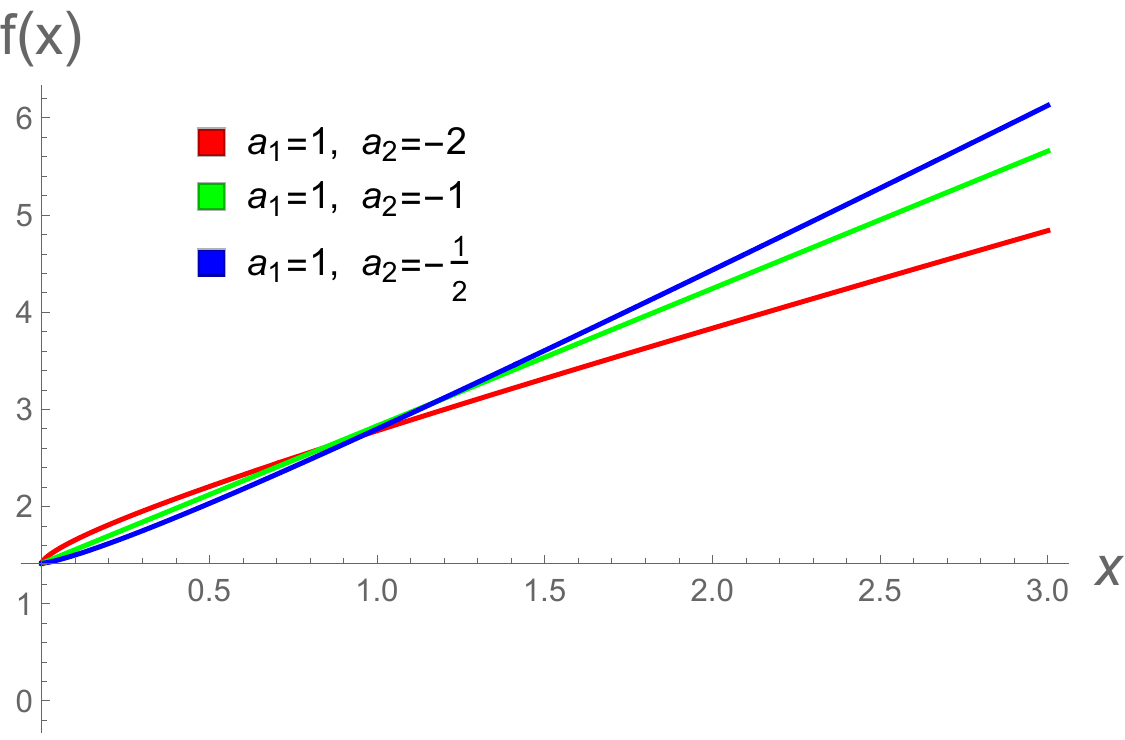}
	\caption{\small In the extremal case $\mu=0$, the function $f(x)$ gives the mass/charge relation, where $M=Q_1 f(x)$ and $x=Q_2/Q_1$. We see that the function is linear if $-a_1 a_2=1$, concave if $-a_1a_2 >1$ and convex when $-a_1a_2<1$.}
	\label{extremalmasscharge}
\end{figure}

We now consider another special case, namely the extremal limit $\mu=0$. The homogeneity conditions of $(M,\Sigma)$ imply that they can be expressed as
\be
M=Q_1 f(x)\,,\qquad \Sigma = Q_1 g(x)\,,\qquad x = \fft{Q_2}{Q_1}\,.
\ee
The long-range force equation \eqref{longrange} and scalar-hair equation \eqref{scalarhaireq} respectively become
\be
4 + 4 x^2 - f^2 - 4g^2=0\,,\qquad a f - 2 g - (a_1-a_2) x f'=0\,,\label{extremalfg}
\ee
There is one integration constant from the above equations and we can fix it based on the discussion of section \ref{sec:emmdbh}. (We cannot use the Schwarzschild black hole as the boundary condition for extremal black holes.) The boundary condition for fixing the integration constant is
\be
f(x_0) = 2 \sqrt{1-\frac{a_1}{a_2}}\,,\qquad x_0=\sqrt{\fft{a_1}{-a_2}}\,.
\ee
With this boundary condition, the equations of \eqref{extremalfg} completely determine the mass/charge and scalar hair/charge relations. For special values of $(a_1,a_2)$, analytical solutions can be obtained. For example, when $a_2=-1/a_1$, we have
\be
f=\fft{2(1+a_1 x)}{\sqrt{a_1^2 + 1}}\,,\qquad
g=\fft{a_1-x}{\sqrt{a_1^2 + 1}}\,.
\ee
When $a_1=\sqrt3=-a_2$, we have
\be
f=\left(x^{2/3}+1\right)^{3/2}\,,\qquad g=-\frac{\sqrt3}{2} \left(x^{2/3}-1\right) \sqrt{x^{2/3}+1}\,.
\ee
These reproduce precisely the known results reviewed in the previous section.  For general $(a_1,a_2)$, we do not have analytical solutions. However, we can obtain some general features by numerical analysis. Firstly, we find that
\be
x\le x_0:\quad g= -\ft12 \sqrt{4+4x^2 - f}\,;\qquad
x\ge x_0:\quad g=+\ft12 \sqrt{4+4x^2-f}\,.
\ee
Secondly, as illustrated in Fig.~\ref{extremalmasscharge}, we find that
\bea
-a_1 a_2>1:&& \hbox{$f(x)$ is concave}\qquad \rightarrow \qquad M(Q_1+Q_2)> M(Q_1) + M(Q_2)\,,\nn\\
-a_1 a_2 = 1:&& \hbox{$f(x)$ is linear}\,\,\,\,\qquad \rightarrow \qquad M(Q_1+Q_2) = M(Q_1) + M(Q_2)\,,\nn\\
-a_1 a_2 <1:&& \hbox{$f(x)$ is convex}\,\qquad  \rightarrow \qquad M(Q_1+Q_2) < M(Q_1) + M(Q_2)\,.\label{sublinsup}
\eea
when $a_1=-a_2$, the equations \eqref{extremalfg} were analysed in \cite{Cremonini:2023vwf, Cremonini:2024eog} in the context of extremal dyonic black holes in EMD theories.

\subsection{Case 3: general $(Q_1,Q_2,\mu)$}

For the general parameters, the homogeneity equations of $M$ and $\Sigma$ imply that we can write them as
\be
M=\mu f(x,y)\,,\qquad \Sigma = \mu g(x,y)\,;\qquad x=\fft{Q_1}{\mu}\,,\qquad y=\fft{Q_2}{\mu}\,.
\ee
The long-range force and scalar-hair equations become respectively the following
\be
1 + 4 x^2 + 4 y^2 - f^2 - 4 g^2 = 0\,,\qquad
a_1 x f_{,x} + a_2 y f_{,y} -2(x g_{,x} +y g_{,y}) =0\,.\label{generalfg}
\ee
For generic values of $(a_1,a_2)$, the partial differential equations cannot be solved exactly. However, as we have already seen, there are two special cases where exact black hole solutions exist. For $a_1 a_2=-1$, we have
\bea
f &=& \frac{\sqrt{4 \left(a_1^2+1\right) x^2+1}}{a_1^2+1} + \frac{\sqrt{4 \left(a_1^{-2}+1\right) y^2+1}}{a_1^{-2}+1}\,,\nn\\
g &=& \frac{a_1 \sqrt{4 \left(a_1^2+1\right) x^2+1}}{2 \left(a_1^2+1\right)} -
\frac{a_1^{-1} \sqrt{4 \left(a_1^{-2}+1\right) y^2+1}}{2 \left(a_1^{-2}+1\right)}\,.
\eea
When $a_1=-a_2=\sqrt3$, the situation is more complicated and we have
\bea
f &=& \fft{1}{2\sqrt3} \left(\sqrt{\frac{\left(\sqrt[3]{u}+1\right) \left(\sqrt[3]{u}+12 x^2+1\right)}{\sqrt[3]{u}}} + \sqrt{\frac{\left(\sqrt[3]{v}+1\right) \left(\sqrt[3]{v}+12 y^2+1\right)}{\sqrt[3]{v}}}\right),\nn\\
g&=&\fft14 \left(\sqrt{\frac{\left(\sqrt[3]{u}+1\right) \left(\sqrt[3]{u}+12 x^2+1\right)}{\sqrt[3]{u}}} - \sqrt{\frac{\left(\sqrt[3]{v}+1\right) \left(\sqrt[3]{v}+12 y^2+1\right)}{\sqrt[3]{v}}}\right)\,,
\eea
where
\bea
u &=& 864 x^4 y^2-18 x^2-1 +6 \sqrt{3} \sqrt{x^4 \left(6912 x^4 y^4-16 x^2 \left(18 y^2+1\right)-16 y^2-1\right)}\,,\nn\\
v &=& 864 x^2 y^4 -18 y^2-1 +6 \sqrt{3} \sqrt{y^4 \left(6912 x^4 y^4-16 x^2 \left(18 y^2+1\right)-16 y^2-1\right)}\,.
\eea
For general $(a_1,a_2)$, we can solve \eqref{generalfg} for $f$ and $g$ numerically. It is necessary to set the boundary conditions. Since the exact solutions of black holes with a single charge, either $x=0$ or $y=0$, exist, we can use these data for the boundary conditions, namely (see case 1.)
\bea
f(x,0) &=& \frac{\sqrt{4 \left(a_1^2+1\right) x^2+1}+a_1^2}{a_1^2+1}\,,\qquad
g(x,0)=\frac{a_1 \left(\sqrt{4 \left(a_1^2+1\right) x^2+1}-1\right)}{2 \left(a_1^2+1\right)}\,,\nn\\
f(0,y) &=& \frac{\sqrt{4 \left(a_2^2+1\right) y^2+1}+a_2^2}{a_2^2+1}\,,\qquad
g(0,y) = \frac{a_2 \left(\sqrt{4 \left(a_2^2+1\right) y^2+1}-1\right)}{2 \left(a_2^2+1\right)}\,.
\eea
It should be pointed out that the partial differential equation in \eqref{generalfg} is singular when $x=0$ or $y=0$. In the numerical analysis, we cannot simply set the boundary of $x$ and $y$ to zero, but sufficiently close to zero for desired accuracy. We can also improve the accuracy by performing a Taylor expansion, such as $f(x,y)=f_0(x) + f_2(x) y^2 + \cdots$ near $y=0$ and the analogous computation near $x=0$. However, for our purpose, such a level of numerical accuracy is unnecessary. One important test of numerical results is that the function $g(x,y)$ must vanish when $y/x = \sqrt{-a_1/a_2}$, in which case, as we discussed in section 2, the solution reduces to the RN black hole.

\begin{figure}
	\centering
	\includegraphics[width=0.4\linewidth]{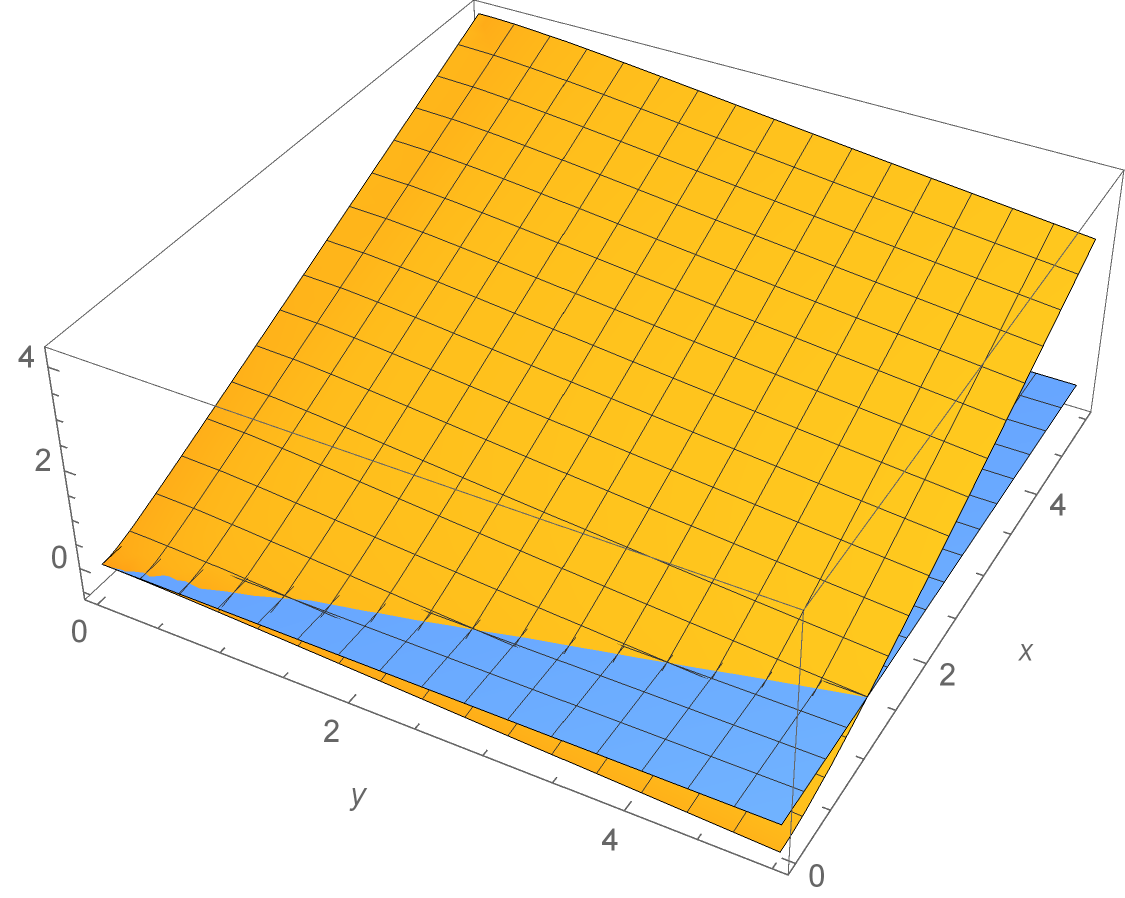} \includegraphics[width=0.4\linewidth]{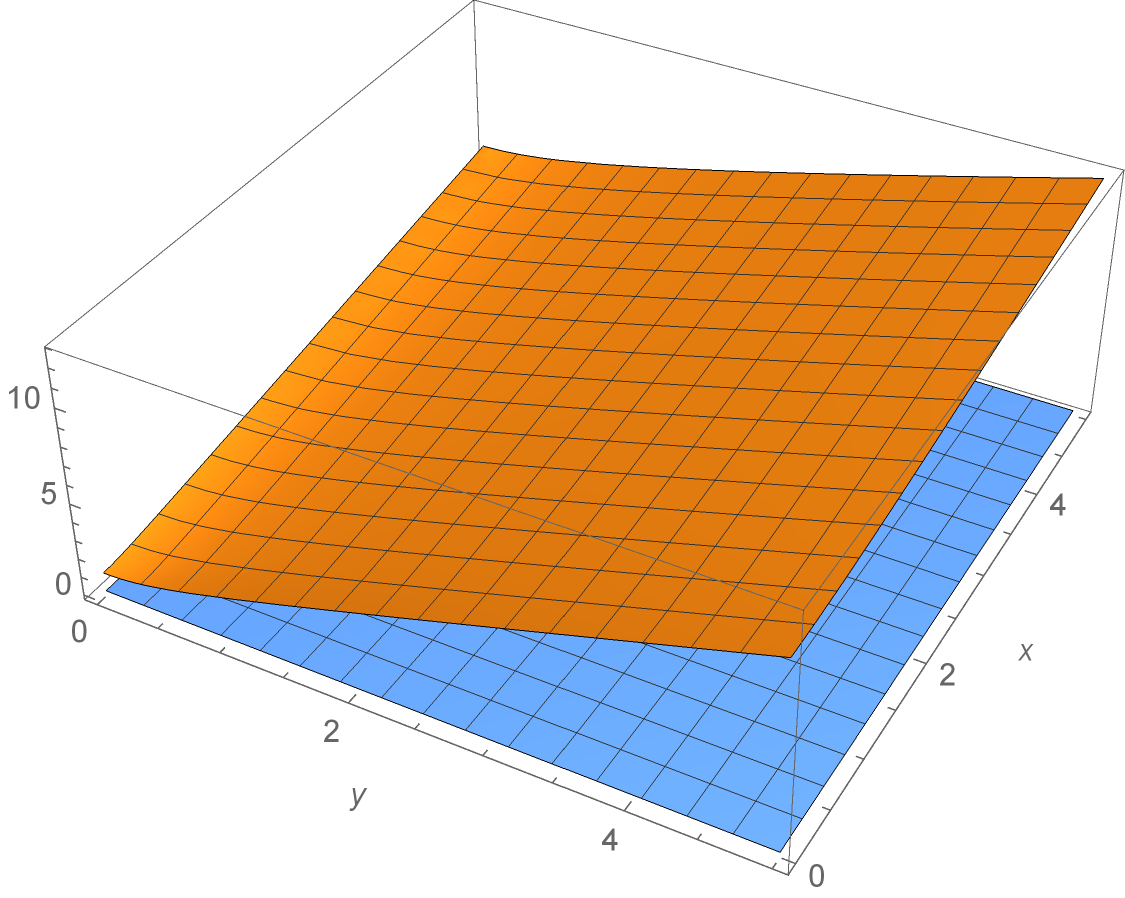}
	\caption{\small Here are the two 3d plots of scalar charge and mass functions $g(x,y)$ and $f(x,y)$ respectively, for the case of $a_1=2$ and $a_2=-1/8$. The flat plane on the bottom in each graph corresponds to the zero value. We see approximately that the scalar charge $g(x,y)=0$ when $y=4 x$, while mass function $f(x,y)$ is always positive.}
	\label{3dplots}
\end{figure}

\begin{figure}
	\centering
	\includegraphics[width=0.6\linewidth]{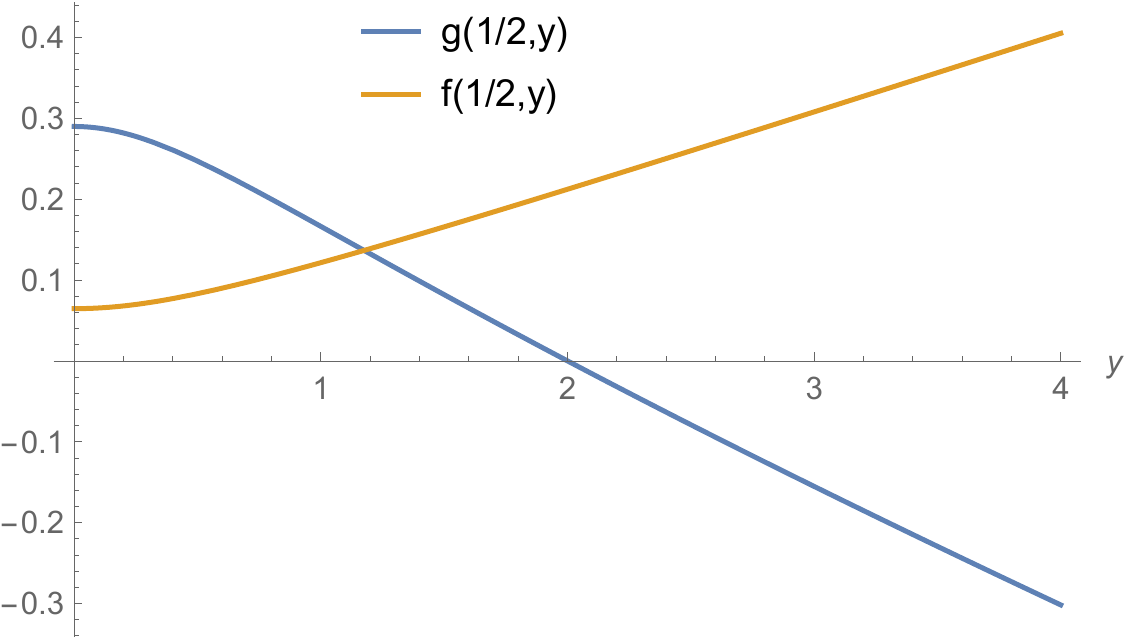}
	\caption{\small In this plot of functions $(f,g)$ with $a_1=2$ and $a_2=-1/8$, we have fixed $x=1/2$. We see that the function $g$ vanishes precisely at $y=2$, where the scalar decouples and the black hole solution becomes the RN black hole.}
	\label{fgyplot}
\end{figure}

As a concrete example for illustration, we consider the case of $a_1=2$ and $a_2=-1/8$. We can use the numerical method to solve for the functions $f(x,y)$ and $g(x,y)$. We plot the 3d graphs for both functions in Fig.~\ref{3dplots}. The 3d plots illustrate that we can obtain both functions $(f,g)$ numerically. The plots are not particularly enlightening in their own right. In Fig.~\ref{fgyplot}, we present a 2D plot for fixed $x=1/2$. Based on the analysis in section \ref{sec:emmdbh}, the black hole becomes the RN black hole when $y=\sqrt{-a_1/a_2}\, x= 4 x$. We see clearly that for $x=1/2$, the function $g$, associated with the scalar charge, indeed vanishes at $y=2$.

\begin{figure}
	\centering
	\includegraphics[width=0.6\linewidth]{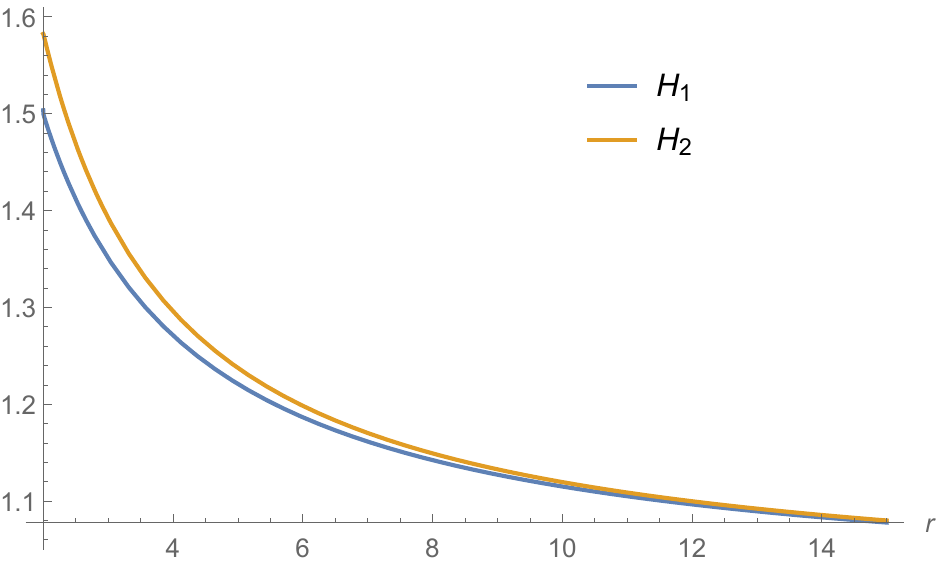}
	\caption{\small For the asymptotic parameters \eqref{m1n1m2} determined by the differential equations of mass and charges, we find that they indeed give rise to a black hole with both $(H_1,H_2)$ monotonically decreasing from finite values at $r=2\mu$ to 1 at large $r$. The graphs are based on $(a_1,a_2)=(2,-1/8)$. The mass and electric charges are $(M,Q_1,Q_2)=(2.42677,1/2,1)$. Consequently, the scalar charge is $\Sigma=0.166408$.}
	\label{h1h2function}
\end{figure}

We now present a concrete generic example with explicit numerical numbers to illustrate our approach. We choose $x=1/2$ and $y=1$, in which case, $(Q_1,Q_2)=(1/2,1)$, if we set $\mu=1$.  The numerical analysis yields
\be
g=0.166408=\Sigma\,,\qquad f=2.42677=M\,.
\ee
Consequently, the parameters of the asymptotic expansion are completely fixed, given by
\be
m_1= 0.240547\,,\qquad n_1=1.18623\,,\qquad m_2 = -0.00779264\,.\label{m1n1m2}
\ee
This set of data indeed gives rise to a black hole solution, as we can see in Fig.~\ref{h1h2function}, where the functions $(H_1,H_2)$ decrease monotonically from some finite values to 1 as the radial coordinate $r$ runs from the horizon $r_+=2\mu$ to asymptotic infinity.

\section{Conclusions}

In this paper, we studied the mass/charge relations of electrically-charged black holes of EMMD theory, inspired by the low-energy effective theories of strings. General spherically-symmetric and static solutions that are asymptotic to the Minkowski spacetime are specified by four parameters: the mass $M$, two electric charges $(Q_1,Q_2)$ and one scalar charge $\Sigma$. Our analysis indicates that the black holes satisfy the weak no-hair theorem involving one real scalar such that the scalar charge $\Sigma$ is not a continuous independent parameter, but a function of the mass and two electric charges, i.e., $\Sigma=\Sigma(M, Q_1,Q_2)$. To find this relation, we proposed a set of algebraic and differential equations that enabled us to determine the relation without having to solve the black hole equations of motion. In particular, in the extremal limit, the mass/charge relation $M(Q_1,Q_2)$ satisfies the sub, linear or super additivity, depending on the values of $a_1 a_2$, as given in \eqref{sublinsup}. We do not have the general proof of the validity of these constraining equations, but in special cases where exact solutions exist, we reproduce the correct mass/charge relations. In the more general cases, we can test them against the numerical results.

It should be pointed out that the weak no-hair theorem of black holes involving only one real scalar only states that the scalar charge is specified by the mass and Maxwell charges, with no additional free continuous parameter. Black holes with fixed mass and Maxwell charges do not in general have to be unique. However, the non-uniqueness spectrum must be discrete since there is no further continuous parameter. A counterexample would invalidate this weak no-hair theorem conjecture. When multiple scalars or further non-gauge fields are involved, continuous hairy parameter is possible, but the weak no-hair theorem conjecture still holds, in that solutions with the most general hair parameters are not black holes, which require further fine-tuning of these parameters.

At first sight, our universal long-range force formula \eqref{longrange} appears to be closely related to the convenient choice of solution ansatz \eqref{ansatz}. One might draw the conclusion that we would need at least to find a clever ansatz and derive the equations of motion in order to find this long-range force condition. Then our claim that black hole mass/charge relation can be obtained without solving equations of motion would be less genuine. However, a clever ansatz is in fact unnecessary, since the parameter $\mu$ can be viewed simply as being introduced to describe strength of the long-range force. This reasoning allows us to generalize the discussion from EMMD theory to more general matter systems with multiple Maxwell fields and dilaton scalars.

For example, we consider Einstein gravities minimally coupled to $n$ Maxwell fields and $m$ dilaton scalars in four dimensions. The Lagrangian takes the form
\be
{\cal L}=\sqrt{-g} \Big(R - \fft12(\partial \vec \phi)^2 - \fft14 \sum_{i=1}^n e^{\vec a_i \cdot \vec \phi} (F_i)^2\Big)\,,\qquad  F_i = dA_i\,,\qquad \vec \phi=(\phi_1,\phi_2,\cdots,\phi_m)\,.
\ee
Note that we require $m\le n$, since when $m\ge n+1$, we can truncate out the scalar components in $\vec \phi$ that are perpendicular to all the dilaton vectors $\vec a_i$.  In view of the weak no-hair theorem, for electrically-charged spherically-symmetric and static black holes that are asymptotic to the Minkowski spacetime, the solutions carry mass $M$, $n$ electric charges $Q_i$ \eqref{Qcharges} only. The $m$ number of scalar charges, defined as $\vec \phi = \fft{4\vec \Sigma}{r} + {\cal O}(1/r^2)$ are not independent. In order to specify the mass and charge relations, we may define a parameter $\mu$ that describes the long-range net force, namely
\be
\lim_{r\rightarrow \infty} r^2\, \hbox{Force} = 4\sum_{i=1}^n Q_i^2 - M^2 - 4 \vec \Sigma \cdot \vec \Sigma = -\mu^2\,.
\ee
With this new parameter, we can write the black hole mass $M$ and the scalar charges $\vec \Sigma$ as homogeneous functions of $(Q_i,\mu)$, i.e.,
\be
M=\sum_{i=1}^n Q_i \fft{\partial M}{\partial Q_i} + \mu \fft{\partial M}{\partial \mu}\,,\qquad
\vec \Sigma=\sum_{i=1}^n Q_i \fft{\partial \vec \Sigma}{\partial Q_i} + \mu \fft{\partial \vec \Sigma}{\partial \mu}\,.
\ee
We conjecture that the relations between the black hole mass $M$ and charges $(Q_i,\vec \Sigma)$ are related by
\be
\vec \Sigma = \fft12 \sum_{i=1}^n \vec a_i Q_i \fft{\partial M}{\partial Q_i} + \mu \fft{\partial \vec \Sigma}{\partial \mu}\,.
\ee
Since exact black hole solutions exist for all the singly-charged cases, we can use these as the boundary conditions for the partial differential equations. We can therefore determine the general mass/charge relation and fix the scalar charges $\vec \Sigma=\vec \Sigma(M, Q_i)$ completely, without having to solve the black hole solutions. It is of great interest to give a full proof that these all indeed give rise to black holes, and to further investigate whether the method of this paper can be extended to study other scalar hairy black holes in higher dimensions and also in general supergravities such as the full STU model and beyond.

Finally we note that the intriguing relation \eqref{tsmu} relates the asymptotic long-range force parameter $\mu$ to the horizon data $TS$. This tells immediately that the long-range force vanishes at the extremal limit. It provides the motivation for studying whether we can determine the full horizon data as functions of asymptotic parameters without having to solve the field equations.

\section*{Acknowledgement}

We are grateful to Zhao Peng for useful discussions. We are also grateful to Hyat Huang and Jie Ren for useful discussions and references. H.L.~is grateful to Peng Huanwu Center for Fundamental Theory, Hefei for hospitality during the finishing stage of this work. This work is supported in part by the National Natural Science Foundation of China (NSFC) grants No.~12375052 and No.~11935009, and also by the Tianjin University Self-Innovation Fund Extreme Basic Research Project Grant No.~2025XJ21-0007. H.L.~is also supported in part by the National Natural Science Foundation of China (NSFC) grants No.12247103.


\begin{thebibliography}{99}

\bibitem{Duff:1995sm}
M.J.~Duff, J.T.~Liu and J.~Rahmfeld,
``Four-dimensional string-string-string triality,''
Nucl. Phys. B \textbf{459}, 125-159 (1996)
doi:10.1016/0550-3213(95)00555-2
[arXiv:hep-th/9508094 [hep-th]].

\bibitem{Duff:1994jr}
M.J.~Duff and J.~Rahmfeld,
``Massive string states as extreme black holes,''
Phys. Lett. B \textbf{345}, 441-447 (1995)
doi:10.1016/0370-2693(94)01638-S
[arXiv:hep-th/9406105 [hep-th]].

\bibitem{Gibbons:1994vm}
G.W.~Gibbons, G.T.~Horowitz and P.K.~Townsend,
``Higher dimensional resolution of dilatonic black hole singularities,''
Class. Quant. Grav. \textbf{12}, 297-318 (1995)
doi:10.1088/0264-9381/12/2/004
[arXiv:hep-th/9410073 [hep-th]].

\bibitem{Lu:1995yn}
H.~L\"u and C.N.~Pope,
``P-brane solitons in maximal supergravities,''
Nucl. Phys. B \textbf{465}, 127-156 (1996)
doi:10.1016/0550-3213(96)00048-X
[arXiv:hep-th/9512012 [hep-th]].

\bibitem{Duff:1999gh}
M.J.~Duff and J.T.~Liu,
``Anti-de Sitter black holes in gauged $N = 8$ supergravity,''
Nucl. Phys. B \textbf{554}, 237-253 (1999)
doi:10.1016/S0550-3213(99)00299-0
[arXiv:hep-th/9901149 [hep-th]].

\bibitem{Sabra:1999ux}
W.A.~Sabra,
``Anti-de Sitter BPS black holes in $N=2$ gauged supergravity,''
Phys. Lett. B \textbf{458}, 36-42 (1999)
doi:10.1016/S0370-2693(99)00564-X
[arXiv:hep-th/9903143 [hep-th]].

\bibitem{Cvetic:1996kv}
M.~Cveti\v c and D.~Youm,
``Entropy of nonextreme charged rotating black holes in string theory,''
Phys. Rev. D \textbf{54}, 2612-2620 (1996)
doi:10.1103/PhysRevD.54.2612
[arXiv:hep-th/9603147 [hep-th]].

\bibitem{Chong:2004na}
Z.W.~Chong, M.~Cveti\v c, H.~L\"u and C.N.~Pope,
``Charged rotating black holes in four-dimensional gauged and ungauged supergravities,''
Nucl. Phys. B \textbf{717}, 246-271 (2005)
doi:10.1016/j.nuclphysb.2005.03.034
[arXiv:hep-th/0411045 [hep-th]].

\bibitem{Rahmfeld:1995fm}
J.~Rahmfeld,
``Extremal black holes as bound states,''
Phys. Lett. B \textbf{372}, 198-203 (1996)
doi:10.1016/0370-2693(96)00063-9
[arXiv:hep-th/9512089 [hep-th]].

\bibitem{Gibbons:1987ps}
G.W.~Gibbons and K.i.~Maeda,
``Black holes and membranes in higher dimensional theories with dilaton fields,''
Nucl. Phys. B \textbf{298}, 741-775 (1988)
doi:10.1016/0550-3213(88)90006-5

\bibitem{Gibbons:1985ac}
G.W.~Gibbons and D.L.~Wiltshire,
``Black holes in Kaluza-Klein theory,''
Annals Phys. \textbf{167}, 201-223 (1986)
[erratum: Annals Phys. \textbf{176}, 393 (1987)]
doi:10.1016/S0003-4916(86)80012-4

\bibitem{Poletti:1995yq}
S.J.~Poletti, J.~Twamley and D.L.~Wiltshire,
``Dyonic dilaton black holes,''
Class. Quant. Grav. \textbf{12}, 1753-1770 (1995)
[erratum: Class. Quant. Grav. \textbf{12}, 2355 (1995)]
doi:10.1088/ 0264-9381/12/7/017
[arXiv:hep-th/9502054 [hep-th]].

\bibitem{Geng:2018jck}
W.J.~Geng, B.~Giant, H.~L\"u and C.N.~Pope,
``Mass of dyonic black holes and entropy super-additivity,''
Class. Quant. Grav. \textbf{36}, no.14, 145003 (2019)
doi:10.1088/1361-6382/ab26e8
[arXiv:1811.01981 [hep-th]].

\bibitem{Cribiori:2022cho}
N.~Cribiori, M.~Dierigl, A.~Gnecchi, D.~L\"ust and M.~Scalisi,
``Large and small non-extremal black holes, thermodynamic dualities, and the swampland,''
JHEP \textbf{10}, 093 (2022)
doi:10.1007/JHEP10(2022)093
[arXiv:2202.04657 [hep-th]].

\bibitem{Cremonini:2023vwf}
S.~Cremonini, M.~Cveti\v c, C.N.~Pope and A.~Saha,
``Mass and force relations for Einstein-Maxwell-dilaton black holes,''
Phys. Rev. D \textbf{107}, no.12, 126023 (2023)
doi:10.1103/Phys RevD.107.126023
[arXiv:2304.04791 [hep-th]].

\bibitem{Cremonini:2024eog}
S.~Cremonini, M.~Cveti\v c, C.N.~Pope and A.~Saha,
``Mass and force relations for extremal Einstein-Maxwell-dilaton-axion black holes,''
Phys. Rev. D \textbf{111}, no.6, 066008 (2025)
doi:10.1103/PhysRevD.111.066008
[arXiv:2412.12277 [hep-th]].

\bibitem{Nelson:2010ig}
W.~Nelson,
``Static Solutions for 4th order gravity,''
Phys. Rev. D \textbf{82}, 104026 (2010)
doi:10.1103/PhysRevD.82.104026
[arXiv:1010.3986 [gr-qc]].

\bibitem{Lu:2015cqa}
H.~L\"u, A.~Perkins, C.N.~Pope and K.S.~Stelle,
``Black holes in higher-derivative gravity,''
Phys. Rev. Lett. \textbf{114}, no.17, 171601 (2015)
doi:10.1103/PhysRevLett.114.171601
[arXiv:1502.01028 [hep-th]].

\bibitem{Huang:2022urr}
Y.~Huang, D.J.~Liu and H.~Zhang,
``Novel black holes in higher derivative gravity,''
JHEP \textbf{02}, 057 (2023)
doi:10.1007/JHEP02(2023)057
[arXiv:2212.13357 [gr-qc]].

\bibitem{Li:2025dpo}
Z.~Li and H.~S.~Liu,
[arXiv:2505.02340 [gr-qc]].

\bibitem{Anabalon:2012ta}
A.~Anabalon,
``Exact Black Holes and Universality in the Backreaction of non-linear Sigma Models with a potential in (A)dS$_4$,'' JHEP \textbf{06}, 127 (2012)
doi:10.1007/JHEP06(2012)127
[arXiv:1204.2720 [hep-th]].

\bibitem{Anabalon:2013qua}
A.~Anabalon, D.~Astefanesei and R.~Mann,
``Exact asymptotically flat charged hairy black holes with a dilaton potential,''
JHEP \textbf{10}, 184 (2013)
doi:10.1007/JHEP10(2013)184
[arXiv:1308.1693 [hep-th]].

\bibitem{Gonzalez:2013aca}
P.A.~Gonz{\'a}lez, E.~Papantonopoulos, J.~Saavedra and Y.~V{\'a}squez,
``Four-dimensional asymptotically AdS black holes with scalar hair,''
JHEP \textbf{12}, 021 (2013)
doi:10.1007/ JHEP12(2013)021
[arXiv:1309.2161 [gr-qc]].

\bibitem{Feng:2013tza}
X.H.~Feng, H.~L\"u and Q.~Wen,
``Scalar hairy black holes in general dimensions,''
Phys. Rev. D \textbf{89}, no.4, 044014 (2014)
doi:10.1103/PhysRevD.89.044014
[arXiv:1312.5374 [hep-th]].

\bibitem{Garfinkle:1990qj}
D.~Garfinkle, G.T.~Horowitz and A.~Strominger,
``Charged black holes in string theory,''
Phys. Rev. D \textbf{43}, 3140 (1991)
[erratum: Phys. Rev. D \textbf{45}, 3888 (1992)]
doi:10.1103/PhysRevD.43.3140

\bibitem{Doneva:2017bvd}
D.D.~Doneva and S.S.~Yazadjiev,
``New Gauss-Bonnet black holes with curvature-induced scalarization in extended scalar-tensor theories,''
Phys. Rev. Lett. \textbf{120}, no.13, 131103 (2018)
doi:10.1103/PhysRevLett.120.131103
[arXiv:1711.01187 [gr-qc]].

\bibitem{Antoniou:2017acq}
G.~Antoniou, A.~Bakopoulos and P.~Kanti,
``Evasion of no-hair theorems and novel black-hole solutions in Gauss-Bonnet theories,''
Phys. Rev. Lett. \textbf{120}, no.13, 131102 (2018)
doi:10.1103/PhysRevLett.120.131102
[arXiv:1711.03390 [hep-th]].

\bibitem{Silva:2017uqg}
H.O.~Silva, J.~Sakstein, L.~Gualtieri, T.P.~Sotiriou and E.~Berti,
``Spontaneous scalarization of black holes and compact stars from a Gauss-Bonnet coupling,''
Phys. Rev. Lett. \textbf{120}, no.13, 131104 (2018)
doi:10.1103/PhysRevLett.120.131104
[arXiv:1711.02080 [gr-qc]].

\bibitem{Cunha:2019dwb}
P.~Cunha, V.P., C.A.R.~Herdeiro and E.~Radu,
``Spontaneously scalarized Kerr black holes in extended scalar-tensor-Gauss-Bonnet gravity,''
Phys. Rev. Lett. \textbf{123}, no.1, 011101 (2019)
doi:10.1103/PhysRevLett.123.011101
[arXiv:1904.09997 [gr-qc]].

\bibitem{Liu:2020yqa}
H.S.~Liu, H.~L\"u, Z.Y.~Tang and B.~Wang,
``Black hole scalarization in Gauss-Bonnet extended Starobinsky gravity,''
Phys. Rev. D \textbf{103}, no.8, 084043 (2021)
doi:10.1103/Phys RevD.103.084043
[arXiv:2004.14395 [gr-qc]].

\bibitem{Herdeiro:2015waa}
C.A.R.~Herdeiro and E.~Radu,
``Asymptotically flat black holes with scalar hair: a review,''
Int. J. Mod. Phys. D \textbf{24}, no.09, 1542014 (2015)
doi:10.1142/S0218271815420146
[arXiv:1504.08209 [gr-qc]].

\bibitem{Herdeiro:2014goa}
C.A.R.~Herdeiro and E.~Radu,
``Kerr black holes with scalar hair,''
Phys. Rev. Lett. \textbf{112}, 221101 (2014)
doi:10.1103/PhysRevLett.112.221101
[arXiv:1403.2757 [gr-qc]].

\bibitem{Zou:2020rlv}
D.C.~Zou and Y.S.~Myung,
``Black hole with primary scalar hair in Einstein-Weyl-Maxwell-conformal scalar theory,''
Phys. Rev. D \textbf{101}, no.8, 084021 (2020)
doi:10.1103/ PhysRevD.101.084021
[arXiv:2001.01351 [gr-qc]].

\bibitem{Lu:1996hh}
H.~L\"u, C.N.~Pope and K.W.~Xu,
``Liouville and Toda solutions of M theory,''
Mod. Phys. Lett. A \textbf{11}, 1785-1796 (1996)
doi:10.1142/S0217732396001776
[arXiv:hep-th/9604058 [hep-th]].

\bibitem{Cremmer:1997xj}
E.~Cremmer, H.~L\"u, C.N.~Pope and K.S.~Stelle,
``Spectrum generating symmetries for BPS solitons,''
Nucl. Phys. B \textbf{520}, 132-156 (1998)
doi:10.1016/S0550-3213(98)00057-1
[arXiv:hep-th/9707207 [hep-th]].

\bibitem{Pacilio:2018gom}
C.~Pacilio,
``Scalar charge of black holes in Einstein-Maxwell-dilaton theory,''
Phys. Rev. D \textbf{98}, no.6, 064055 (2018)
doi:10.1103/PhysRevD.98.064055
[arXiv:1806.10238 [gr-qc]].

\bibitem{Herdeiro:2025blx}
C.~Herdeiro, E.~Radu and E.~dos Santos Costa Filho,
``Charged, rotating black holes in Einstein-Maxwell-dilaton theory,''
[arXiv:2506.15798 [gr-qc]].

\bibitem{Duff:1996hp}
M.J.~Duff, H.~L\"u, C.N.~Pope, ``The black branes of M-theory,''
Phys. Lett. B \textbf{382}, 73-80 (1996)
doi:10.1201/9781482268737-29
[arXiv:hep-th/9604052 [hep-th]].

\bibitem{Cvetic:1996gq}
M.~Cveti\v c and A.A.~Tseytlin,
``Nonextreme black holes from nonextreme intersecting M-branes,''
Nucl. Phys. B \textbf{478}, 181-198 (1996)
doi:10.1016/0550-3213(96)00411-7
[arXiv:hep-th/9606033 [hep-th]].

\bibitem{Cremonini:2022sxf}
S.~Cremonini, M.~Cveti\v c, C.N.~Pope and A.~Saha,
``Long-range forces between nonidentical black holes with non-BPS extremal limits,''
Phys. Rev. D \textbf{106}, no.8, 086007 (2022)
doi:10.1103/PhysRevD.106.086007
[arXiv:2207.00609 [hep-th]].


\bibitem{Lu:2013eoa}
H.~L\"u,
``Charged dilatonic AdS black holes and magnetic AdS$_{D-2} \times R^{2}$ vacua,''
JHEP \textbf{09}, 112 (2013)
doi:10.1007/JHEP09(2013)112
[arXiv:1306.2386 [hep-th]].

\bibitem{Lu:2013uia}
H.~L{\"u} and W.~Yang,
``$SL(n,R)$-Toda black holes,''
Class. Quant. Grav. \textbf{30}, 235021 (2013)
doi:10.1088/0264-9381/30/23/235021
[arXiv:1307.2305 [hep-th]].

\end{thebibliography}
\end{document}